\documentclass{article}
\pdfoutput=1
\usepackage{amsmath,amsfonts,amssymb,graphicx,color,amsbsy,bm,gensymb,calrsfs,arydshln,float,authblk,hyperref}

\DeclareMathAlphabet{\pazocal}{OMS}{zplm}{m}{n}

\newcommand{\bu}{\mathbf{u}}
\newcommand{\bv}{\mathbf{v}}
\newcommand{\bx}{\mathbf{x}}
\newcommand{\bnu}{\boldsymbol{\nu}}
\DeclareMathOperator*{\argmin}{\text{argmin}}

\title{Variational approach to contact line dynamics \\ for thin films}

\author{Dirk Peschka}
\affil{Weierstrass Institute for Applied Analysis and Stochastics \\ Mohrenstr. 39, 10117 Berlin, Germany\\ e-mail: peschka@wias-berlin.de}
\date{\today}

\begin{document}
\maketitle
\begin{abstract}
This paper investigates a variational approach to viscous flows with contact line dynamics based on energy-dissipation modeling. The corresponding model is reduced to a thin-film equation and its variational structure is also constructed and discussed. Feasibility of this modeling approach is shown by constructing a numerical scheme in 1D and by computing numerical solutions for the problem of gravity driven droplets. Some implications of the contact line model are highlighted in this setting.
\end{abstract}

\section{Introduction}

The wetting and dewetting flow of a thin layer of a viscous liquid over a solid planar surface has supplied researchers with a valuable model system to study a number of interesting  problems \cite{de1985wetting,oron1997long,bonn2009wetting}. Related phenomena appear in nature, but are also of great importance for applications, \emph{e.g.}, droplet splashing, wetting, coating, painting, pattern formation processes, multiphase flows, and microfluidics, to name only a few.
The mathematical and numerical analysis of the corresponding free boundary problem is considered quite challenging.
Moving contact lines create a classical singularity that needs to be resolved and different mechanisms for this have been proposed \cite{huh1971hydrodynamic,hocking1976moving,shikhmurzaev1993moving,lauga2007microfluidics}.
Such models allow contact lines to move and to relax towards an equilibrium.
However, usually one observes dynamic contact angles and phenomena related to advancing and receding angles or even hysteretic contact line motion. For an introduction and an overview of different approaches for contact line models,  including variational approaches, we refer to \cite{dussan1979spreading,snoeijer2013moving,qian2006variational,ren2007boundary,sui2014numerical}.

The general idea of a gradient flow is to construct an abstract state space $q\in Q$ and a corresponding vector space of velocities $\dot{q}\in V$. The evolution of states $q(t)$ is then driven by an energy $\mathcal{E}:Q\to\mathbb{R}$. In the context of physical systems, this energy could be a thermodynamic potential for systems with diffusion and heat transport or it could be a potential energy for mechanical systems. Additionally, the construction requires a dissipation functional $\mathcal{D}:Q\times V\to\mathbb{R}$, which in many cases is non-negative and quadratic in the second argument and depends on the state. The dissipation $\mathcal{D}$ operates similar to a Riemanian metric and allows to define gradients $\nabla_\mathcal{D}\mathcal{E}\in V$  by
$\langle {\rm D}_{\dot{q}} \mathcal{D}(q,\nabla_\mathcal{D}\mathcal{E}),\dot{p}\rangle = \langle {\rm D}_q \mathcal{E}(q),\dot{p}\rangle$ for all $\dot{p}\in V$.
The corresponding gradient flow then solves
\begin{align}
\dot{q}(t)=-\nabla_\mathcal{D}\mathcal{E}\bigl(q(t)\bigr),
\end{align}
with decreasing energy $\tfrac{{\rm d}}{{\rm d}t}\mathcal{E}\bigl(q(t)\bigr) = \langle {\rm D}_q \mathcal{E},\dot{q}\rangle=-\langle {\rm D}_{\dot{q}} \mathcal{D}(q,\dot{q}),\dot{q}\rangle \le 0$ by construction. We deliberately employ the dot notation to indicate both membership
$\dot{p}\in V$ and time-derivatives $\dot{q}=\tfrac{\partial}{{\partial}t} q$.
The gradient flow is mathematically equivalent to the minimization problem
\begin{align}
\dot{q}=\argmin_{\dot{p}\in V}\Big(\frac{1}{2}\mathcal{D}(q,\dot{p})+\langle {\rm D}_q\mathcal{E},\dot{p}\rangle\Big),
\end{align}
which is a useful statement when considering alternative approaches and when adding constraints to the state space $Q$ and to the velocity space $V$, see \emph{e.g.} Peletier~\cite{peletier2014variational}.
The gradient flow construction usually applies to the irreversible dynamics of systems where driving forces are entirely balanced by friction. Other examples of variational structures are Poisson or symplectic structures, which are suitable for reversible processes and conserve energy. In the context of fluid flow the Euler equation has a Poisson structure, whereas the Stokes equation is dissipative.

The goal of this paper is to support the development of models for moving contact lines by the formal construction of a variational gradient flow model, by establishing efficient numerical algorithms, and by further exploring modeling ideas. Even though this work contains some useful ideas for the Navier-Stokes or Stokes equation, the primary focus of this work is to translate these concepts to thin-film models.
Therefore, in Section \ref{sec:geomodel} we introduce the Stokes gradient flow construction with free boundaries and provide the necessary definitions for the problem geometry, for the energy and for the dissipation of viscous Newtonian fluids.
The highlight in this construction is a dissipation term at the contact line, which creates a particular model relating contact line velocity and contact angle, \emph{i.e.}, a contact line model.
We show that the Stokes flow with free boundaries and contact line model can be recovered using a Helmholtz-Rayleigh dissipation principle, \emph{i.e.}, a gradient flow, using the so-called flow-map as the state variable. Then, in Section \ref{sec:thinfilm} we perform a thin-film reduction, which is based on scaling arguments in the energy and the dissipation. Particular care is taken in treating all the terms at boundaries and contact lines correctly.
We also derive a variational formulation of the thin-film model, which includes dissipation in bulk (viscosity), at interfaces (Navier-slip), and at contact lines (contact line dissipation).
We discuss the numerical implementation of this model in detail in one spatial dimension. The generalization to higher dimensions is straight-forward.
Finally, we present a couple of examples showing gravity driven moving droplets, where we highlight the effect of contact line dissipation and perform some robustness tests with the proposed numerical algorithm.
For the clarity of presentation some helpful calculations
and definitions have been moved to the Appendix~\ref{sec:appendix}.

\section{Stokes gradient flow}%
\label{sec:geomodel}
\subsection{Geometry, flows, and functionals}
\begin{figure}[t]
\centering
\includegraphics[width=0.5\textwidth]{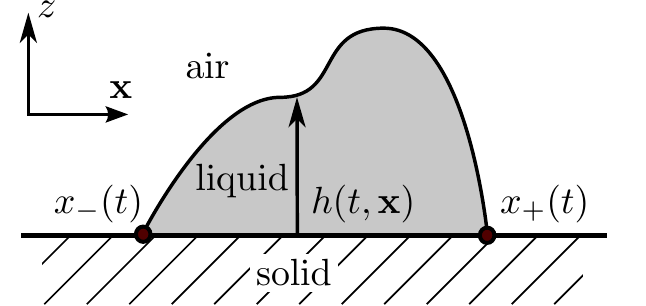}
\caption{Sketch of viscous liquid on a planar solid surface parametrized by $h$ }
\label{fig:geom}
\end{figure}
The motion of a viscous liquid layer occupying the domain
\begin{subequations}
\label{eqn:geometry}
\begin{align}
\Omega(t)=\{(\bx,z)\in\mathbb{R}^{d+1}:0<z<h(t,\bx)\},
\end{align}
can often be described using a non-negative time-dependent function $h(t,\bx)\ge 0$, which para\-metrizes the height of the liquid-air interface over the solid surface at $z=0$ as shown in Fig.~\ref{fig:geom}. Furthermore, we denote the support of the function $h$ with
\begin{align}
\omega(t)=\{\bx\in\mathbb{R}^d:h(t,\bx)>0\},
\end{align}
\end{subequations}
and the contact line is the set $\gamma(t)=(\partial\omega(t),0)$. Due to its prominent appearance  $\bnu\in\mathbb{R}^d$ is used for the outer normal to $\omega$ on $\partial\omega$. Otherwise, the notation $\bnu_S$ refers to the outer normal to $S$ on $\partial S$.
Later on, we consider the special case $d=1$ and  $\omega$ is connected, so the support is a single interval
written as $\omega(t)=\bigl(x_-(t),x_+(t)\bigr)$ for $x_\pm(t)\in\mathbb{R}$ and correspondingly the contact line consists of the two points $\gamma(t)=\{(x_-(t),0\bigr),\bigl(x_+(t),0)\}$.
In order to emphasize the relation of these sets and the state-space $Q$ one usually uses the concept of flow maps
\begin{align}
q(t)=\bigl(\mathbf{X}(t,\cdot),Z(t,\cdot)\bigr):\Omega_0\to\Omega,
\end{align}
that are homeomorphisms between $\Omega_0=\Omega(0)$ and admissible shapes $\Omega=\Omega(t)$ at time $t$.
Thereby $q$ also maps any of the introduced domains at $t=0$ to their shape at time $t>0$. On the level of coordinates  $\mathbf{y}\in\Omega_0$ we write $\bx=\mathbf{X}(t,\mathbf{y})$, $z=Z(t,\mathbf{y})$ with  $(\bx,z)\in\Omega$. For fluids the corresponding velocity field
$\dot{q}(t)=\bigl(\dot{\mathbf{X}}(t,\cdot),\dot{Z}(t,\cdot)\bigr):\Omega_0\to\mathbb{R}^{d+1}$
is often expressed in Eulerian coordinates $\dot{q}\simeq \bu=(\bu_\bx,u_z):\Omega\to\mathbb{R}^{d+1}$ by composing $\dot{q}$ with the inverse map $q^{-1}$.

In this sense, the Stokes or Navier-Stokes equation for the
flow field $\bu(t,\bx,z)$ can be understood as an evolution equation for the flow map $q$. For incompressible liquids this flow field is divergence free $\nabla\cdot\bu=0$.
For the moment  assume that the evolution is driven purely by a surface energy of the form
\begin{align}
\label{eqn:energy}
\mathcal{E}(t)=\theta_\ell |\Gamma_\ell(t)|+\theta_\text{s$\ell$}|\Gamma_\text{s$\ell$}(t)|,
\end{align}
where $\Gamma_{\ell}(t)=\{(\bx,z):z=h(t,\bx)>0\}$ denotes the free liquid-air interface, the free solid-liquid interface is $\Gamma_{\text{s$\ell$}}(t)=(\omega(t),0)$, with $|S|$ we denote the Lebesgue measure of the set $S$, and
the surface tension coefficients are $\theta_\ell=\theta_\text{liquid,air}>0$ and $\theta_\text{s$\ell$}=\theta_\text{solid,liquid}-\theta_\text{solid,air}$.
By introducing the flow map the energy depends on the state $\mathcal{E}(t)\equiv\mathcal{E}\bigl(q(t)\bigr)$. Using $(h,\omega)$ to represent the flow map, one can rewrite the surface measures $|\Gamma_\ell|$ and $|\Gamma_{\text{s$\ell$}}|$ explicitly as
\begin{subequations}
\label{eqn:interfaces}
\begin{align}
\label{eqn:interfacesA}
&|\Gamma_{\ell}|(q)=\int_{\Gamma_\ell}{\rm d}s=\int_{\omega} \sqrt{1+|\nabla h|^2}\,{\rm d}\bx,\\
\label{eqn:interfacesB}
&|\Gamma_{\text{s$\ell$}}|(q)=\int_{\Gamma_\text{s$\ell$}}{\rm d}s=\int_{\omega}{\rm d}\bx
=|\omega|\overset{d=1}{=}x_+-x_-.
\end{align}
\end{subequations}
Throughout this paper $\nabla$ and $\nabla\cdot$ denote the gradient and divergence in both $d+1$ and in $d$ dimensions acting on all available spatial arguments, \emph{i.e.}, on $(\bx,z)$ or on $\bx$ as it should be clear from the context.
For the model derivation it is instructive to write all relevant friction mechanism of this system. First, the dissipation for the bulk velocity is
\begin{subequations}
\label{eqn:dissipation}
\begin{align}
\label{eqn:diss_bulk}
\mathcal{D}_\Omega(\bu)=\int_\Omega \boldsymbol{\tau}(\bu):\nabla\bu\,\,{\rm d}\bx\,{\rm d}z,
\end{align}
where for Newtonian fluids the shear stress is of the form $\boldsymbol{\tau}(\bu)=2\mu_\Omega\mathbb{D}(\bu)$ with liquid viscosity $\mu_\Omega>0$ and the symmetric gradient $\mathbb{D}(\bu)=\tfrac{1}{2}(\nabla\boldsymbol{\bu}+\nabla\boldsymbol{\bu}^\top)$.
Moving contact lines are known to produce logarithmic singularities for no-slip conditions, \emph{i.e.}, when we would simply set $\bu=0$ on $\Gamma_\text{s$\ell$}$. Instead, we just require impermeability $\bu\cdot\mathbf{e}_z\equiv u_z=0$ on $\Gamma_\text{s$\ell$}$ and introduce the dissipation at the solid-liquid interface
\begin{align}
\label{eqn:diss_interface}
\mathcal{D}_\omega(\bu)=\int_{\Gamma_\text{s$\ell$}} {\mu_\omega}\bu_\bx^2\,{\rm d}s,
\end{align}
where $\mu_\omega\ge 0$ is related to the well-known Navier-slip length $b$ via $\mu_\omega=\mu_\Omega/b$. Consequently, we also introduce a quadratic dissipation mechanism at the contact line $\gamma$, which reads
\begin{align}
\label{eqn:diss_cl}
\mathcal{D}_\gamma(\bu)=\int_{\gamma} {\mu_\gamma}\bu_\bx^2\,{\rm d}\gamma={\mu_\gamma}\big(\dot{x}^2_-+\dot{x}^2_+\big),
\end{align}
where the latter reformulation is only meaningful for $d=1$ and $\dot{x}_\pm=\frac{d}{dt}x_\pm$ denotes the contact line velocity and $\mu_\gamma\ge 0$ is the friction coefficient. For higher dimensions $d>1$ it makes sense to further decompose the contact line dissipation according to
$\mathcal{D}_\gamma(\bu)=\int_\gamma \mu_{\perp} (\bu_\bx\cdot\bnu)^2 + \mu_{\parallel} ([1-\bnu\bnu^\top]\bu_\bx)^2\,{\rm d}\gamma$.
When all dissipation terms are collected, the total dissipation is
\begin{align}
\label{eqn:diss_total}
\mathcal{D}(\bu)=\mathcal{D}_\Omega(\bu)+\mathcal{D}_\omega(\bu)+\mathcal{D}_\gamma(\bu),
\end{align}
\end{subequations}
which is a positive quadratic functional for incompressible flow fields $\bu$. The obvious dependence of $\mathcal{D}:Q\times V\to\mathbb{R}$ on the state $q$ through the shape of $\Omega,\omega,\gamma$ is not stated explicitly as an argument. The evolution of the domain, which can be represented by $(h,\omega)$, is restricted by the constraints of incompressibility 
and by the kinematic condition
\begin{align}
\label{eqn:kinematics}
\dot{h} + \nabla \cdot \left(\int_0^h \bu_\bx\,{\rm d}z\right)=0.
\end{align}
This allows to assign velocities
$(\dot{h},\dot{\bx})$ to $(h,\omega)$, where $\dot{\bx}:\gamma\to\mathbb{R}^d$ is the velocity of $\gamma$ obtained by restricting $\bu_\bx$ to $\gamma$. While the states are still the flow-maps, the concept of shapes parametrized with $(h,\omega)$ will be helpful later.

\subsection{Gradient flow construction}
\begin{figure}[t]
\centering
\includegraphics[width=0.5\textwidth]{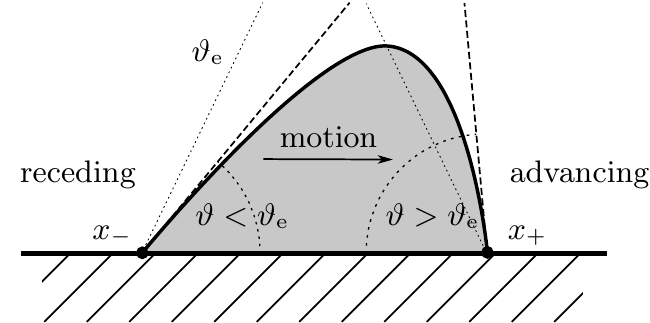} 
\caption{Sketch of contact lines $x_\pm$ and angles $\vartheta$ and of advancing and receding contact angles for an exemplary liquid droplet sliding to the right}
\label{fig:clm}
\end{figure}

Following the gradient flow recipe, we obtain the Stokes equation for the unknown domain via the constrained minimization problem 
\begin{align}
\label{eqn:min_diss_energy}
\bu(t,\cdot)=\argmin_{\bv\in V}\Big(\frac{1}{2}\mathcal{D}(\bv)+\langle {\rm D}_q\mathcal{E},\bv\rangle\Big),
\end{align}
for which differentiation at $\bu$ in direction of $\bv$ results in the weak formulation $a(\bu,\bv)=f(\bv)$ for all $\bv\in V$. The bilinear form $a(\bu,\bv):=\tfrac{1}{2}\langle{\rm D}_{\dot{q}}\mathcal{D}(\bu),\bv\rangle$ is
\begin{subequations}
\label{eqn:stokes_weakform}
\begin{align}
a(\bu,\bv)=&\int_\Omega {\mu_\Omega}\,\mathbb{D}(\bu){\,:\,}\mathbb{D}(\bv)\,{\rm d}\bx\,{\rm d}z + 
\int_{\Gamma_\text{s$\ell$}} \mu_\omega \bu_\bx\cdot\bv_\bx\,{\rm d}s + 
\int_\gamma \mu_\gamma\,\dot{\bx}\cdot\bv_\bx\,{\rm d}\gamma,
\end{align}
and the linear functional is $f(\bv):=\langle -{\rm D}_q \mathcal{E},\bv\rangle$, which using the tangential gradient $\bar{\nabla}$ can be written
\begin{align}
\langle {\rm D}_q \mathcal{E},\bv\rangle = \,&\theta_\ell\int_{\Gamma_\ell} \bar{\nabla} {\rm id_\Gamma}:\bar{\nabla} \bv\,{\rm d}s + 
\theta_\text{s$\ell$}\int_{\Gamma_\text{s$\ell$}} \bar{\nabla} {\rm id_\Gamma}:\bar{\nabla} \bv\,{\rm d}s.
\end{align}
\end{subequations}
In order to (formally) reconstruct the strong form of the differential equation, we use integration by parts on curved surfaces
\begin{align*}
\int_\Gamma \bar{\nabla} {\rm id_\Gamma}{:}\bar{\nabla} \bv\,{\rm d}s = -d\int_\Gamma \kappa \bnu_\Omega\cdot\bv\,{\rm d}s + \int_{\partial\Gamma}\bv\cdot\bnu_\Gamma\,{\rm d}\gamma,
\end{align*}
where $\kappa$ denotes the signed mean curvature (see Appendix~\ref{appendix:hints}), ${\rm id_\Gamma}$ is the coordinate identity on the surface, $\bnu_\Omega$ is the outer normal of $\Omega$ on the free surface $\Gamma\subset\partial\Omega$,
 $\bnu_\Gamma$ is the outer normal of $\Gamma$ on $\partial\Gamma$, and finally ${\rm d}s$ and ${\rm d}\gamma$ are the integration measures of $\Gamma$ and $\partial\Gamma$.
Note that due to impermeability $\bv\cdot\mathbf{e}_z=0$ only the term at
$\partial\Gamma_\text{s$\ell$}$ contributes from ${\rm D}_q|\Gamma_\text{s$\ell$}|$, so that in total we have
\begin{align}
\nonumber\langle {\rm D}_q \mathcal{E},\bv\rangle = -d\theta_\ell\int_{\Gamma_\ell} \kappa\bnu_\Omega\cdot\bv\,{\rm d}s  + 
\int_\gamma
\mathbf{f}_\gamma\cdot\bv\,{\rm d}\gamma,
\end{align}
with the force assigned to the contact line is defined as
$\mathbf{f}_\gamma=(\theta_\ell\bnu_{\Gamma_\ell} +
 \theta_\text{s$\ell$}\bnu_{\Gamma_\text{s$\ell$}})$. Sometimes $\textbf{f}_\gamma$ is referred to as uncompensated Young force (see Appendix~\ref{appendix:young}). In total this produces the strong form of the PDE for the unknown velocity $\bu(t,\cdot):\Omega(t)\to\mathbb{R}^{d+1}$ so that
\begin{subequations}
\label{eqn:stokes}
\begin{align}
-\nabla p  + \nabla\cdot\boldsymbol{\tau}(\bu)=0\qquad\qquad &\text{in }\Omega(t),\\[0.1cm]
\nabla\cdot\bu=0\qquad\qquad &\text{in }\Omega(t),\\[0.2cm]
\mathbf{t}\cdot(\boldsymbol{\tau}\cdot \bnu_\Omega + \mu_\omega\bu)=0\qquad\qquad &\text{on } \Gamma_\text{s$\ell$}(t),\\[0.1cm]
(-p\mathbb{I}+\boldsymbol{\tau})\cdot \bnu_\Omega=d\theta_\ell\kappa\bnu_\Omega\qquad\qquad&\text{on }\Gamma_\ell,(t)\\[0.2cm]
\label{eqn:clm}(\dot{\bx}, u_z)^\top=\mu_\gamma^{-1}(\mathbb{I}-\mathbf{e}_z\mathbf{e}_z^\top)\,\mathbf{f}_\gamma \qquad\qquad&\text{at }\gamma(t),
\end{align}
\end{subequations}
where the pressure $p(t,\cdot):\Omega(t)\to\mathbb{R}$ is added as a Lagrange multiplier to account for the incompressibility in the constrained minimization \eqref{eqn:min_diss_energy}.
Then, the domain $\Omega(t)$ evolves according to the kinematic condition \eqref{eqn:kinematics}.
The first mathematical analysis of well-posedness of such a problem (without contact lines) is by Beale \cite{beale1981initial}. The discretization of such a model and in particular the discretization of the curvature using finite elements was discussed by B\"ansch  \cite{bansch2001finite}. Note that \eqref{eqn:clm} in the Stokes equation is a \emph{contact line model} with \emph{receding} and \emph{advancing} contact angle terminology as sketched in Fig.~\ref{fig:clm} and corresponds to the continuum model proposed by Ren \& E \cite{ren2007boundary}. For an introduction to variational modeling and gradient flows, and in particular the mathematical equivalence of different energetic variational principles, we refer to the lecture notes by Peletier~\cite{peletier2014variational}.
\section{Thin-film models}
\label{sec:thinfilm}
\subsection{Model dimension reduction}
Now we discuss, how the free boundary problem \eqref{eqn:stokes} can be reduced to a thin-film type model for the film height $h$ and its support set $\omega$. Particular care will be taken in the treatment of boundary and interface terms. A cornerstone of the lubrication model is the non-dimensionalization of length and velocity scales via 
\begin{align*}
\bx=L\tilde{\bx}, \quad z=\varepsilon L\tilde{z},\quad \bu_\bx = U\tilde{\bu}_\bx, \quad u_z =  \varepsilon U \tilde{u}_z,
\end{align*}
for typical length $L$ and typical velocity $U=L/T$ with a small parameter $0<\varepsilon\ll 1$. If we expand the solution $\tilde{\bu}$ in an asymptotic series in $\varepsilon$ and expand the weak form to leading order we get
\begin{align*}
\frac{a(\bu,\bv)}{U^2L^{d-1}\mu_\Omega}=&\frac{1}{\varepsilon}\int_\Omega (\partial_{\tilde{z}} \tilde{\bu}_\bx)(\partial_{\tilde{z}} \tilde{\bv}_\bx)\,\,{\rm d}\tilde{\bx}\,{\rm d}\tilde{z}
\quad + \\ &\quad
\frac{L}{b}\int_{\Gamma_\text{s$\ell$}} \tilde{\bu}_\bx\cdot\tilde{\bv}_\bx\,\,{\rm d}\tilde{s}
\quad +  \\
&\qquad
\frac{\mu_\gamma}{\mu_\Omega}\int_\gamma \tilde{\bu}_\bx\cdot\tilde{\bv}_\bx\,\,{\rm d}\tilde{\gamma}
\quad + \quad
 \text{l.o.t.},
\end{align*}
where all terms contribute to the leading order if $b\sim\varepsilon L$ and $\mu_\gamma \sim \varepsilon^{-1}\mu_\Omega$ as $\varepsilon\to 0$. Therefore, we denote by $\tilde{b}=b/(\varepsilon L)$ and  $\tilde{\mu}_\gamma=\varepsilon\tilde{\mu}_\gamma/\mu_\Omega$ the rescaled non-dimensional parameters. The remaining lower order terms (l.o.t.) are omitted for brevity.
Having $\tilde{b}\sim 1$ is realistic in some microfluidic settings since  slip-lengths between nanometers and micrometers have been observed experimentally. When one is
interested macroscopic dynamic of thin films, the contact line singularity might only be resolved at a microscopic length scale $0<L_\text{m}\ll \varepsilon L$, which then leads to an apparent contact line friction  $\mu_\gamma \sim \mu_\Omega \log(\varepsilon L/L_\text{m})$. However, $\mu_\gamma$ might very well be an intrinsic physical parameter on its own right. For the moment, we consider the energy contribution from \eqref{eqn:interfacesA} and perform the thin-film reduction by expanding
\begin{align*}
|\Gamma_\ell|
=\int_\omega \sqrt{1+|\nabla h|^2}\,{\rm d}\bx
=L^d\int_\omega \Big( 1 + \frac{\varepsilon^2}{2}|\tilde{\nabla} \tilde{h}|^2 \Big)  \,{\rm d}\tilde{\bx}+\text{l.o.t.},
\end{align*}
so that the derivative of this functional is
\begin{align*}
\frac{\langle {\rm D}_q|\Gamma_\ell|,\bv\rangle}{L^{d-1}U}=&\int_\omega \varepsilon^2\tilde{\nabla} \tilde{h}\cdot\tilde{\nabla} \dot{\tilde{h}}_\bv  \,{\rm d}\tilde{\bx}+
\int_{\gamma}\Big(1+\frac{\varepsilon^2}{2}|\tilde{\nabla} \tilde{h}|^2\Big) \tilde{\bv}_\bx\cdot\bnu\,{\rm d}\tilde{\gamma} +
\text{l.o.t.},
\end{align*}
where we associate $\dot{h}_\bv + \nabla\cdot\int_0^h \bv_\bx\,{\rm d}z =0$ to any $\bv\in V$ as in \eqref{eqn:kinematics} and $\tilde{\nabla}$ acts on $\tilde{\bx}$. In order to arrive at this derivative one needs to use Reynold's transport theorem, to take into account the derivative with respect to the motion of the support. Using the definition of $\dot{h}_\bv$ we can rewrite the first term as
\begin{align*}
\int_\omega \tilde{\nabla} \tilde{h}\cdot\tilde{\nabla} \dot{\tilde{h}}_\bv \,{\rm d}\tilde{\bx}=-\int_\omega (\tilde{\Delta} \tilde{h}) \dot{\tilde{h}}_\bv \,{\rm d}\tilde{\bx} + \int_{\gamma} \dot{\tilde{h}}_\bv\tilde{\nabla}_{\bnu} \tilde{h}\,{\rm d}\tilde{\gamma}\\
=\int_\omega \tilde{\Delta} \tilde{h}\left[\tilde{\nabla}\cdot\int_0^h \tilde{\bv}_\bx\,{\rm d}\tilde{z}\right] \,{\rm d}\tilde{\bx}+ \int_{\gamma} \dot{\tilde{h}}_\bv\tilde{\nabla}_{\bnu} \tilde{h}\,{\rm d}\tilde{\gamma}\\
=\int_\Omega -\tilde{\nabla}\tilde{\Delta} \tilde{h}\cdot \tilde{\bv}_\bx\, {\rm d}\tilde{\bx}\,{\rm d}\tilde{z}+ \int_{\gamma} \dot{\tilde{h}}_\bv\tilde{\nabla}_{\bnu} \tilde{h}\,{\rm d}\tilde{\gamma},
\end{align*}
where the last term on the boundary in the integration-by-parts vanished, because $\int_0^h\tilde{\bv}_\bx\,{\rm d}\tilde{z}\equiv 0$ on $\gamma$.
Since $h\equiv 0$ on $\gamma$ we have $\dot{h}_\bv+\bv_\bx\cdot\nabla h=0$ for the convective derivative. This allows to transform the last term into
\begin{align*}
\int_{\gamma} \dot{\tilde{h}}_\bv\tilde{\nabla}_{\bnu} \tilde{h}\,{\rm d}\tilde{\gamma} = \int_\gamma -\tilde{\bv}_\bx\cdot\bnu|\tilde{\nabla}\tilde{h}^2|\,{\rm d}\tilde{\gamma}.
\end{align*}
For each of the surfaces we get the derivative to be
\begin{align*}
\frac{\langle {\rm D}_q|\Gamma_\ell|,\bv\rangle}{L^{d-1}U}=&\int_\Omega -\varepsilon^2\tilde{\nabla}\tilde{\Delta} \tilde{h}\cdot \tilde{\bv}_\bx\, {\rm d}\tilde{\bx}\,{\rm d}\tilde{z}+
\int_{\gamma}(1-\tfrac{\varepsilon^2}{2}|\tilde{\nabla} \tilde{h}|^2) \tilde{\bv}_\bx\cdot\bnu\,{\rm d}\tilde{\gamma} + 
\text{l.o.t.,}\\
\frac{\langle {\rm D}_q|\Gamma_\text{s$\ell$}|,\bv\rangle}{L^{d-1}U}=&\int_{\gamma}\tilde{\bv}_\bx\cdot\bnu\,{\rm d}\tilde{\gamma},
\end{align*}
so that for the derivative of the energy we get
\begin{align*}
  \frac{\langle {\rm D}_q\mathcal{E},\bv\rangle}{L^{d-1}U}=&\int_\Omega -\varepsilon^2\theta_\ell\tilde{\nabla}\tilde{\Delta} \tilde{h}\cdot \tilde{\bv}_\bx\, {\rm d}\tilde{\bx}\,{\rm d}\tilde{z}\qquad +\\
  & \qquad \int_{\gamma}\Big[\theta_\text{s$\ell$}+\theta_\ell(1-\tfrac{\varepsilon^2}{2}|\tilde{\nabla} \tilde{h}|^2)\Big] \tilde{\bv}_\bx\cdot\bnu\,{\rm d}\tilde{\gamma}+\text{l.o.t.}
\end{align*}
For all terms to contribute at the same order, we require $\theta_\ell+\theta_\text{s$\ell$}\sim \varepsilon^2$. This lets us define $\varepsilon$ so that  $\varepsilon^2=(\theta_\ell+\theta_\text{s$\ell$})/\theta_\ell$ is indeed small.
Finally, in order to balance dissipation and energy we define the velocity scale
\begin{align}
U=\varepsilon^3\frac{\theta_\ell}{\mu_\Omega},
\end{align}
so that only the global length scale $L$ remains to be defined. Putting all contributions from energy and dissipation in one expression and dropping the tilde from all expressions gives
\begin{align*}
\int_\omega\left(\int_0^h -\partial_{zz}\bu_\bx \bv_\bx\,{\rm d}z\right) + \Big(\partial_z\bu_\bx\bv_\bx\Big)^h_0\,{\rm d}\bx+
\int_{\Gamma_\text{s$\ell$}} b^{-1}\bu_\bx\cdot\bv_\bx\,{\rm d}\bx+
\int_{\gamma} \mu_\gamma \,\dot{\bx}\cdot\bv_\bx{\rm d}\gamma=\\
\int_\Omega\nabla\Delta h\cdot\bv_\bx\,{\rm d}\bx\,{\rm d}z-\int_{\gamma}(1-\tfrac{1}{2}|\nabla h|^2)\bv_\bx\cdot\bnu\,{\rm d}\gamma,
\end{align*}
which needs to hold for all test functions $\bv\in V$. We can integrate this equation for $\bu_\bx(t,\bx,z)$ in $z$ and obtain the explicit expression
\begin{align}
  \bu_\bx = -\frac{z^2}{2}\nabla\Delta h + c_1z+c_0,
\end{align}
where $c_0(t,\bx),c_1(t,\bx)$ are determined by the boundary conditions $\partial_z\bu_\bx=0$ at $z=h$ and $b\,\partial_z\bu_\bx + \bu_\bx=0$ at $z=0$ implied by the boundary terms before. At the contact line $\gamma$ one obtains the law
\begin{align}
\label{eqn:contactlinelaw}
\mu_\gamma\,\dot{\bx} = \left(\frac{1}{2}|\nabla h|^2 -1\right)\bnu,
\end{align}
so that the contact line is advancing, receding, or stationary for $|\nabla h|>\sqrt{2}$, $|\nabla h|<\sqrt{2}$, or $|\nabla h|=\sqrt{2}$, respectively. In the thin-film approximation we have contact angles $\vartheta\sim\varepsilon\nabla h$ and the equilibrium contact angle $\vartheta_\text{e}=\varepsilon\sqrt{2}$, so that we have the contact line dynamics in the slightly more familiar form
\begin{align}
\mu_\gamma\,\dot{\bx} = \frac{1}{2\varepsilon^2}\Big(\vartheta^2-\vartheta_\text{e}^2\Big)\bnu.
\end{align}
The last step is to insert the explicit expression for $\bu_\bx$ into the kinematic condition \eqref{eqn:kinematics}. This returns the degenerate, fourth-order parabolic equation
\begin{subequations}
\label{eqn:thinfilm}
\begin{align}
\label{eqn:tf1}
\dot{h} &= \nabla\cdot\Big(m(h)\,\nabla\,(-\Delta h)\Big),\\
\label{eqn:tf2}
\mu_\gamma\,\dot{\bx} &= \left(\frac{1}{2}|\nabla h|^2 -1\right)\bnu,
\end{align}
for the height $h$ and the boundary $\bx$ alone.
Additionally we have the constraint $\dot{h} + \dot{\bx}\cdot\nabla h =0$ on $\gamma$. The mobility $m$ encodes the dissipation mechanism in the bulk and at the interface, where our integration gives
\begin{align}
\label{eqn:mobility}
m(h)=\frac{h^3}{3}+b h^2,
\end{align}
\end{subequations}
where $b$ encodes the rescaled slip-length that we introduced before.

Employing the rescaled parametrization on the original energy $\mathcal{E}$ gives
\begin{align*}
\mathcal{E}&=L^d\int_\omega \theta_\ell\sqrt{1+\varepsilon^2|\nabla h|^2}+\theta_\text{s$\ell$}\,{\rm d}\bx\\
&=L^d\int_\omega \theta_\ell\left(1+\frac{\varepsilon^2}{2}|\nabla h|^2\right)+\theta_\text{s$\ell$}\,{\rm d}\bx+\text{l.o.t.}\\
&\approx\varepsilon^2L^d\theta_\ell\,\int_\omega \left(1 + \
\frac{1}{2}|\nabla h|^2\right)\,{\rm d}\bx=:\varepsilon^2 L^d\theta_\ell\,\,\mathcal{E}_\text{tf},
\end{align*}
where in the last step we used the previous definition of $\varepsilon^2=(\theta_\ell+\theta_\text{s$\ell$})/\theta_\ell$ to define the thin-film energy $\mathcal{E}_\text{tf}$.
Additionally, we are going to add a trivial bulk term $\int_\omega f(h;\bx)\,{\rm d}\bx$ to the thin-film energy $\mathcal{E}_\text{tf}$. Since we assume $f(0;\bx)\equiv 0$, this term does not directly contribute to the contact line law. The variational formulation of the thin-film model is discussed now in more detail.

\subsection{Variational thin-film model}

Let us consider the simpler problem of finding the velocities $\dot{h}\in V$ with $\dot{h}:\omega\to\mathbb{R}$
that solve the a thin-film model in the weak form on a \emph{fixed} domain $\omega$ and strictly positive $h$.
The corresponding variational formulation requires the introduction of an additional variable $\pi:\omega\to\mathbb{R}$, which is related to a given $\dot{h}\in V$ through the degenerate elliptic equation
\begin{align}
\label{eqn:tf_pressure_constraint}
\dot{h}-\nabla\cdot(m(h)\nabla \pi)=0,
\end{align}
with homogeneous natural boundary conditions  $m\bnu\cdot\nabla\pi=0$.
Then, the thin-film model on a fixed domain has a gradient structure with the dissipation
\begin{subequations}
\begin{align}
\label{eqn:tf_dissipation}
\mathcal{D}_\text{tf}(\dot{h}_v)=\int_\omega m(h)|\nabla\pi|^2\,{\rm d}\bx,
\end{align}
and with the driving thin-film energy
\begin{align}
\label{eqn:tf_energy}
\mathcal{E}_\text{tf}(h)=\int_\omega 1+\frac{1}{2}|\nabla h|^2+f(h;\bx)\,  {\rm d}\bx.
\end{align}
\end{subequations}


The evolution of the height $h(t,\bx)$ is again governed by a constrained minimization problem similar to the one in \eqref{eqn:min_diss_energy} stated as
\begin{align}
\label{eqn:min_diss_tfenergy}
\dot{h}(t,\cdot)=\argmin_{\dot{h}_v\in V}\Big(\frac{1}{2}\mathcal{D}_\text{tf}(\dot{h}_v)+\langle {\rm D}_h\mathcal{E}_\text{tf},\dot{h}_v\rangle\Big),
\end{align}
and can be written in the block form
\begin{align}
\label{eqn:tf_blockform}
\begin{pmatrix}
0 & 0 & M^* \\
0 & D & D^* \\
M & D & 0
\end{pmatrix}
\begin{pmatrix}
\dot{h} \\ \pi \\ \lambda
\end{pmatrix}
=
\begin{pmatrix}
  -Sh-M\partial_h f\\ 0 \\ 0
  \end{pmatrix},
\end{align}
where $\lambda:\omega\to\mathbb{R}$ is the Lagrange multiplier enforcing \eqref{eqn:tf_pressure_constraint}
and the operators $M,D,S$ are defined as
\begin{subequations}
\label{eqn:operators}
\begin{align}
\langle v,M w\rangle  &=\int_\omega vw\,{\rm d}\bx,\\
\langle v,D\,w \rangle &=\int_\omega m(h)\nabla v\cdot\nabla w\,{\rm d}\bx, \\
\langle v,S\,w \rangle &=\int_\omega \nabla v\cdot\nabla w \,{\rm d}\bx,
\end{align}
\end{subequations}
and map function spaces for $\dot{h},\pi,\lambda$ into appropriate dual function spaces. We presume the operators $D,M,S$ are self-adjoint.
The first line of the block operator \eqref{eqn:tf_blockform} implies $\lambda=\Delta h-\partial_h f$, whereas the second line implies basically $\pi=-\lambda=-\Delta h=\delta \mathcal{E}_\text{tf}/\delta h$, so that the third line $\dot{h}-\nabla\cdot(m\nabla\pi)=0$ produces the thin-film equation as in \eqref{eqn:tf1}
\begin{align}
  \dot{h}-\nabla\cdot\big(m(h)\nabla\pi\big)=0,\qquad \pi=-\Delta h + \partial_h f(h;\bx).
\end{align}
We introduced the saddle-point problem \eqref{eqn:tf_blockform} in preparation for the more complex gradient structure with contact line dynamics, where additional bulk-interface coupling terms are required.
%
Now we will state the gradient form of the contact line model.

As before, in the presence of a moving contact line, $h$ is supported only on the set $\omega(t)$ with the contact line defined as the set $\gamma(t)=\partial\omega(t)$. Let $\bx(t)\in\gamma(t)$ be a point
on the contact line, then $h(t,\bx(t))\equiv 0$ for all $t$ and consequently
\begin{align}
\label{eqn:tf_velocity_constraint}
\tfrac{{\rm d}}{{\rm d}t}h(t,\bx(t))=\dot{h}(t,\bx(t))+\dot{\bx}(t)\cdot\nabla\, h\bigl(t,\bx(t)\bigr) =0,
\end{align}
Since \eqref{eqn:tf_velocity_constraint} does not depend on the parameterization, it is justified to simply write $\dot{h}+\dot{\bx}\cdot\nabla h=0$ with $\dot{\bx}:\gamma\to\mathbb{R}^d$ the contact line speed introduced before. This condition relates the normal component of
$\dot{\bx}$ with the Eulerian time derivative $\dot{h}$ of the height.

Analogous to the gradient structure of the Stokes model we can now understand $q=(h,\omega)$ as the general state space with velocities $\dot{q}=(\dot{h},\dot{\bx})$. The corresponding dissipation for the thin-film model analogous to \eqref{eqn:tf_dissipation} but with contact line dynamics is
\begin{align}
\mathcal{D}_\text{tf}(\dot{q})=\int_{\omega} m(h)|\nabla \pi|^2\,{\rm d}\bx + \int_\gamma \mu_\gamma \dot{\bx}^2\,{\rm d}\gamma,
\end{align}
and the driving thin-film energy $\mathcal{E}_\text{tf}$ from \eqref{eqn:tf_energy}. Since now the support can change we use Reynolds' transport theorem to include variations of the energy \eqref{eqn:tf_energy} with respect to the shape of $\omega$ via%
\begin{align*}
  \langle {\rm D}_q \mathcal{E}_\text{tf},\dot{q}\rangle=\int_\omega \nabla h \cdot \nabla \dot{h} + \partial_h f(h;\bx)\,{\rm d}\bx
+\int_\gamma \left(1+\tfrac{1}{2}|\nabla h|^2\right)(\dot{\bx}\cdot\bnu) \,{\rm d}\gamma.
\end{align*}
In the standard thin-film model we only have to enforce the constraint $\dot{h}-\nabla\cdot(m\nabla\pi)=0$ in $\omega$ using a multiplier $\lambda:\omega\to\mathbb{R}$, whereas now we additionally have to enforce $\dot{h}+\dot{\bx}\cdot\nabla h=0$ on the contact line $\gamma$ with a multiplier $\kappa:\gamma\to\mathbb{R}$. Additionally, the resulting linear system has a potential ambiguity with respect to adding a constant to $\pi$, which we resolve by adding a constraint
 $\int\pi\,{\rm d}\bx=0$ with a multiplier $\rho\in\mathbb{R}$. Therefore we seek $u=(\dot{h},\pi,\dot{\bx},\lambda,\kappa,\rho)^\top$ solving the corresponding constrained minimization problem leading to $Au=-b$ where $A:W\to W^*$ is defined as
 \begin{subequations}
 \label{eqn:blockform}
 \begin{align}
 \def\arraystretch{1.2}
 Au=\!\left(\begin{array}{ccc:ccc}
 \cdot & \cdot & \cdot & M^* & M_{\gamma}^* & \cdot \\
 \cdot & D & \cdot & D^* & \cdot & M_0^* \\
 \cdot & \cdot & D_\gamma & \cdot & C_\gamma^* & \cdot \\
 \hdashline
 M & D & \cdot           & \cdot & \cdot & \cdot \\
 M_{\gamma} & \cdot & C_\gamma & \cdot & \cdot & \cdot \\
 \cdot & M_0 & \cdot         & \cdot & \cdot & \cdot
 \end{array}\right)\!
 \left(\begin{array}{c}
 \dot{h}\\ \pi \\ \dot{\bx} \\ \hdashline \lambda \\ \kappa \\ \rho
 \end{array}\right),
\end{align}
and $b\in W^*$ is defined as
\begin{align}
\def\arraystretch{1.2}
 b=\!\left(\begin{array}{c}
 \delta_h \mathcal{E}_\text{tf} \\ 0 \\ \delta_\bx \mathcal{E}_\text{tf} \\ \hdashline 0 \\ 0 \\ 0
 \end{array}\right)=\left(\begin{array}{c}
 Sh+M\partial_h f \\ 0 \\ \bnu M_\gamma(1+\tfrac{1}{2}|\nabla h|^2) \\ \hdashline 0 \\ 0 \\ 0
 \end{array}\right)\!.
 \end{align}
 \end{subequations}
\noindent
 Note that $M,D,S$ are defined in \eqref{eqn:operators}, whereas $M_\gamma,C_\gamma,D_\gamma$ act as follows
  \begin{subequations}
 \begin{align}
 \langle \bv,D_\gamma\,\mathbf{w} \rangle&=\int_\gamma \mu_\gamma \bv\cdot\mathbf{w}\,{\rm d}\gamma, \\
 \langle v,M_\gamma w\rangle&=\int_\gamma vw\,{\rm d}\gamma, \\
 \langle v,C_\gamma \mathbf{w}\rangle&=\int_\gamma v\,\mathbf{w}\cdot\nabla h \,{\rm d}\gamma,
 \end{align}
\end{subequations}
where $\bv,\mathbf{w}:\gamma\to\mathbb{R}^d$, $v:\gamma\to\mathbb{R}$ and $w:\omega\to\mathbb{R}$.
Additionally we defined $M_0\,\pi =\int_\omega \pi\,{\rm d}\bx$. The dashed lines in the definition of the matrix $A$ divide the dissipation part (upper left block) and the constraints of the problem (remaining blocks). All the parts of $A$ with a dot $(\cdot)$ are entirely zero. As before, from the block structure of $A$ we can reconstruct the PDE with the contact line model, which we briefly outline. As before, we first eliminate the unknown Lagrange multiplier $\lambda,\kappa,\rho$. The first line of $A$ gives $\lambda=\Delta h - \partial_h f$ and $\kappa=-\bnu\cdot\nabla h$. The third line then gives
\begin{align*}
\mu_\gamma\dot{\bx}&=-\kappa\nabla h -\bnu\Big(1+\frac{1}{2}|\nabla h|^2\Big)=\bnu\left(\frac{1}{2}|\nabla h|^2-1\right),
\end{align*}
where we used $\nabla h = -\bnu|\nabla h|$ and $\bnu\cdot\nabla h=-|\nabla h|$ to arrive at
the contact line model which we already observed directly in \eqref{eqn:tf2}. From the second line we get $\rho=0$ and $\pi=-\lambda$ and by inserting this into the constraint $\dot{h}-\nabla\cdot(m\nabla \pi)=0$ recover the thin-film model in \eqref{eqn:thinfilm}. Note, including $\rho$ removes a potential zero eigenvalue from the algebraic system. This identification confirms that the gradient structure in \eqref{eqn:blockform} corresponds to the thin-film model we obtained by the formal asymptotics.

\begin{figure*}[t]
\centering
\includegraphics[height=0.26\textwidth]{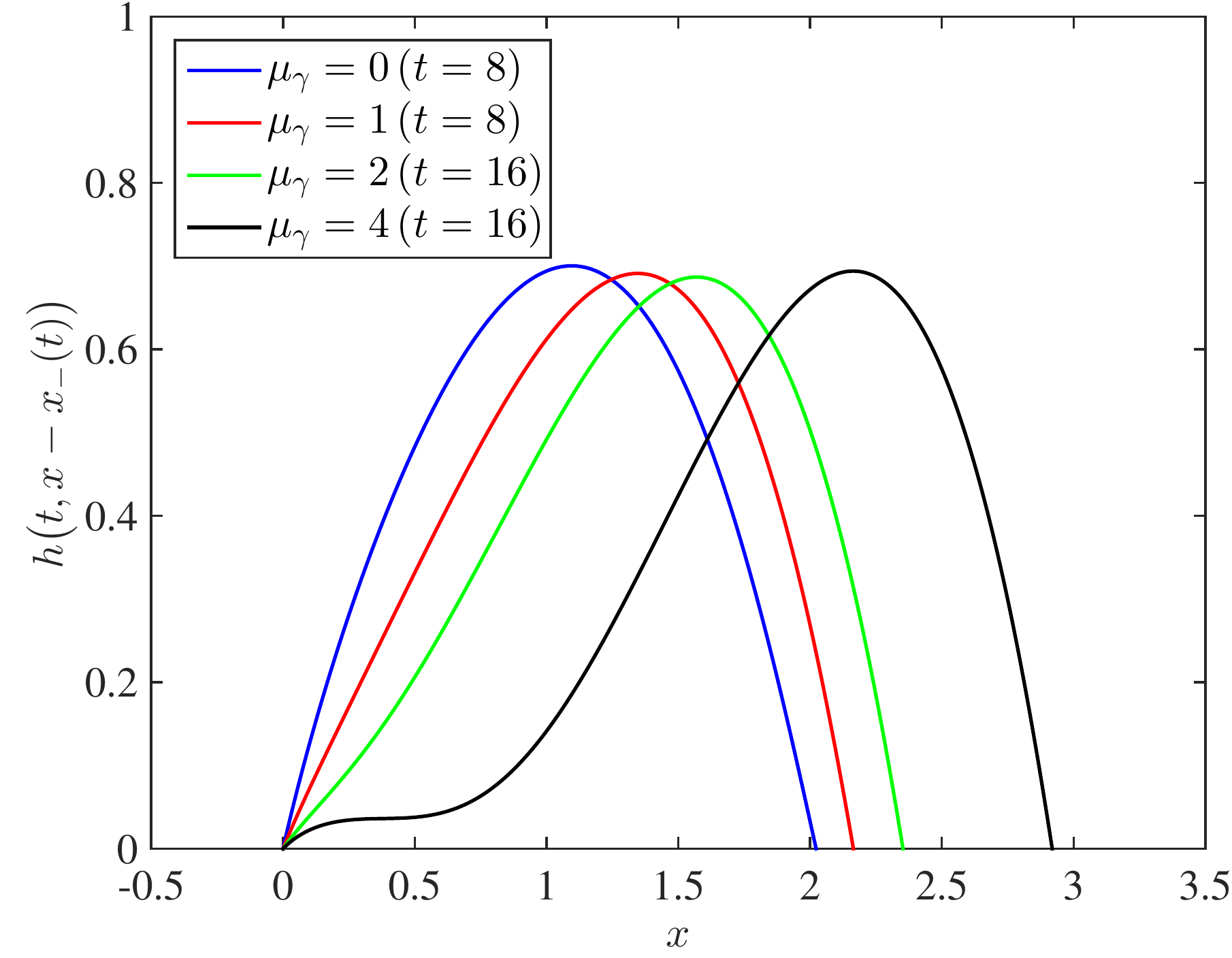}\,%
\includegraphics[height=0.26\textwidth]{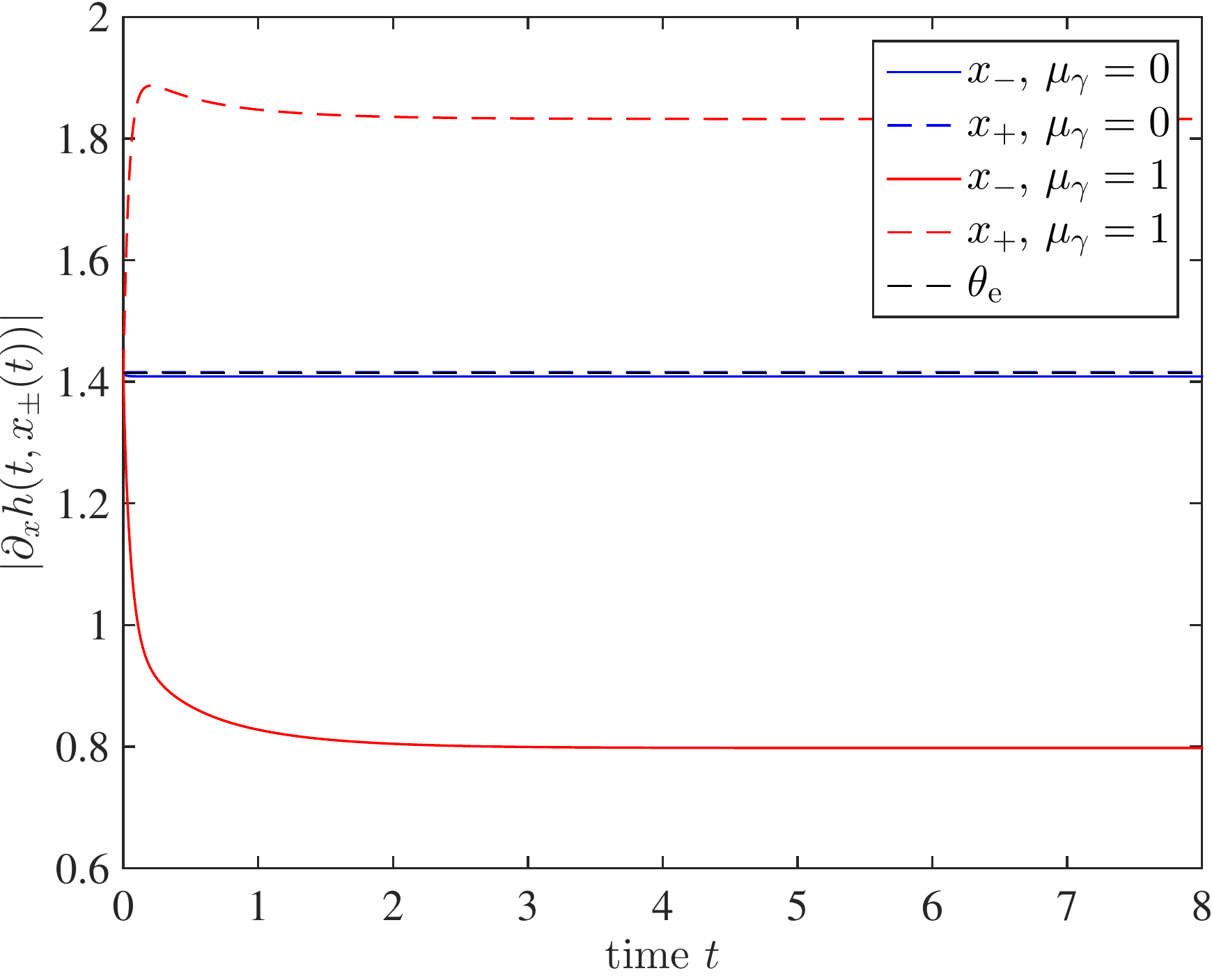}\,%
\includegraphics[height=0.26\textwidth]{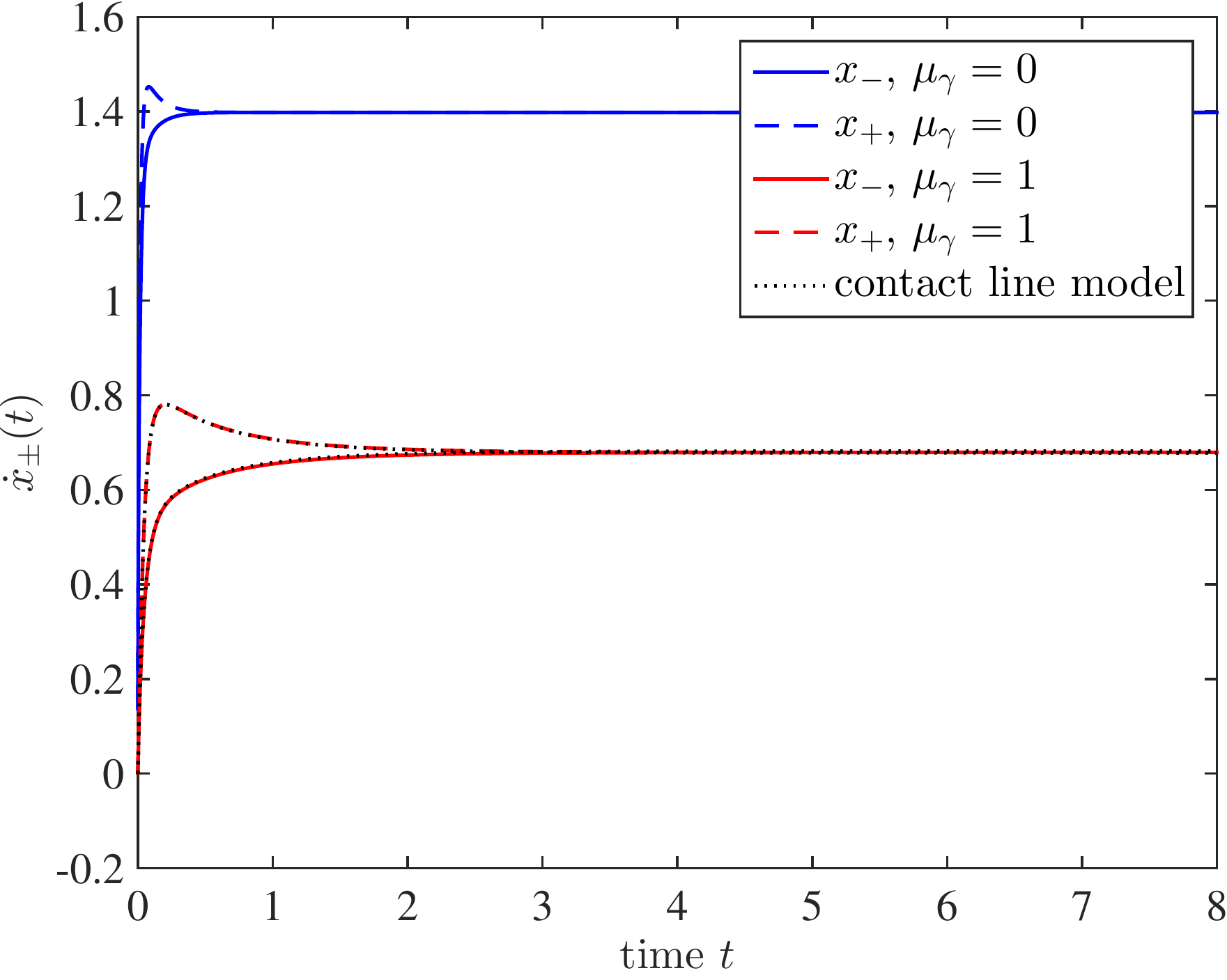}%
\caption{Solutions of the thin-film model with contact line dynamics for $\mu_\gamma=0$ (static angle model) and $\mu_\gamma=1,2,4$ (dynamic angle model) showing \textbf{(left)} the shifted height profile $h(t,x-x_-)$ \textbf{(middle)} contact angle \emph{vs} time and \textbf{(right)} contact line velocity \emph{vs} time}
\label{fig:solution}
\end{figure*}

\subsection{Numerical implementation}
The spatial discretization of weak formulation in \eqref{eqn:blockform} and is performed using standard $P_1$ finite elements. Since the resulting PDE will be of fourth-order parabolic type, an implicit time-discretization is advantageous to overcome restrictions of the time-step size $\tau$ due to a CFL-type condition. A semiimplicit discretization can be achieved by replacing $Sh$ in $b$ with $S(h+\tau\dot{h})$, which effectively modifies $A$ such that we have the component $A_{\dot{h}\dot{h}}=\tau S$. A similar strategy might be useful for $\delta_\bx \mathcal{E}_\text{tf}$, but was not needed so far. Once $\dot{h}$ is computed by solving $Au=-b$,
one can extract $\dot{h}$.
However, it makes absolutely no sense to define $h(t+\tau,\cdot)=h(t,\cdot)+\tau\dot{h}$, since $h(t+\tau,\cdot)$ and $h(t,\cdot)$ are defined on different domains $\omega(t)$ and $\omega(t+\tau)$. However, using a diffeomorphism  $\boldsymbol{\xi}(t,\cdot):\omega(t_0)\to\omega(t)$ as we have defined it in Appendix~\ref{appendix:ALE}, we
can pull-back $H(t,\mathbf{y})=h\bigl(t,\boldsymbol{\xi}(t,\mathbf{y})\bigr)$ to the reference domain
%
and recover time-derivates $(\dot{H},\dot{\boldsymbol{\xi}})$ in the \emph{Arbitrary Lagrangian-Eulerian} (ALE) frame and
%
update
the corresponding discrete solution 
\begin{subequations}
\begin{align}
H(t+\tau,\mathbf{y})&=H(t,\mathbf{y})+\tau\dot{H}(\mathbf{y}),\\
\boldsymbol{\xi}(t+\tau,\mathbf{y})&=\boldsymbol{\xi}(t,\mathbf{y})+\tau\dot{\boldsymbol{\xi}}(\mathbf{y}),
\end{align}
\end{subequations}
in the ALE reference frame. More details for the decomposition $\dot{h}\mapsto (\dot{H},\dot{\boldsymbol{\xi}})$ and  for the construction of mappings in higher spatial dimensions but without contact line dynamics can be found in  \cite{peschka2015thin}. The corresponding 1D MATLAB code \texttt{thinfilm\_clm.m} is available as a GitHub repository \cite{peschkaGitHub}.

\subsection{Gravity driven droplets in $d=1$}

For $L=2$ and $\omega=(0,L)$ we use the initial data
\begin{align*}
h_0(x)=\frac{\sqrt{2}}{L}x(L-x), 
\end{align*}
constituting a $d=1$ droplet with equilibrium contact angles at $x_\pm(0)=\{0,L\}$. For the mobility we use $m(h)=h^2$ and study the evolution for contact line dissipation $\mu_\gamma=\{0,1,2,4\}$. Furthermore, we use the extra contribution to the energy
\begin{align*}
f(h;x)=-ghx,
\end{align*}
to include tangential gravity, where we set $g=3$ to drive liquid volumes in the positive $x$-direction and obtain traveling wave solutions for long times. The problem for $\mu_\gamma=0$ and $\mu_\gamma=1$ is solved for $0\le t\le T=8$, whereas the problem for $\mu_\gamma=2$ and $\mu_\gamma=4$ is solved for $0\le t\le T=16$. In order to study the experimental rate of convergence in space we solve the problem on domains
discretized using $2^{m}+1$ points for $m=1\ldots10$ using $N_\tau=25\,000$ uniform time steps. We then compare solutions at the final time $T$ with respect to the convergence of $x_\pm(T)$ and use an affine map to pull back the solution $h(T,x)$ to a fixed domain $\omega_0\approx\omega(T)$. On the fixed domain we study the convergence of $h$ with respect to the $L^2(\omega_0)$ and the $H^1(\omega_0)$ norm as $\delta x \sim 2^{-m} L \to 0$.

In the left panel of Fig.~\ref{fig:solution} we show the solution of a droplet moving nearly with constant velocity, \emph{i.e.}, nearly a traveling wave solution, for different contact line dissipations $\mu_\gamma$. Evidently, with static contact angles $\vartheta=\vartheta_\text{e}$ the droplet is rather symmetric with respect to reflections, whereas for $\mu_\gamma=1,2,4$ it becomes increasingly asymmetric. This is most noticeable for $\mu_\gamma=4$, where it appears that contact line friction can force droplets to develop a 'nose' and for larger $\mu_\gamma$ possibly undergo a topological transition. Also note the tendency for larger $\mu_\gamma$ that the contact angle at $x_-$, \emph{i.e.}, the receding side, is decreasing, while it is increasing at $x_+$, \emph{i.e.}, the advancing side.

The middle panel of Fig.~\ref{fig:solution} depicts the temporal evolution of the contact angle $|\partial_x h\bigl(t,x_\pm(t)\bigr)|$ for $\mu=0,1$, which for $\mu_\gamma$ is close to the equilibrium value of $\sqrt{2}$, as for $\mu_\gamma=1$ we observe the before mentioned behavior of $0\le \partial_x h(t,x_-) < \sqrt{2} < \bigl(-\partial_x h(t,x_+)\bigr)$. Note that for $t>3$ the contact angles are nearly constant, so that the solution is already quite close to a traveling wave.

\begin{figure*}[t]
\centering
\includegraphics[height=0.26\textwidth]{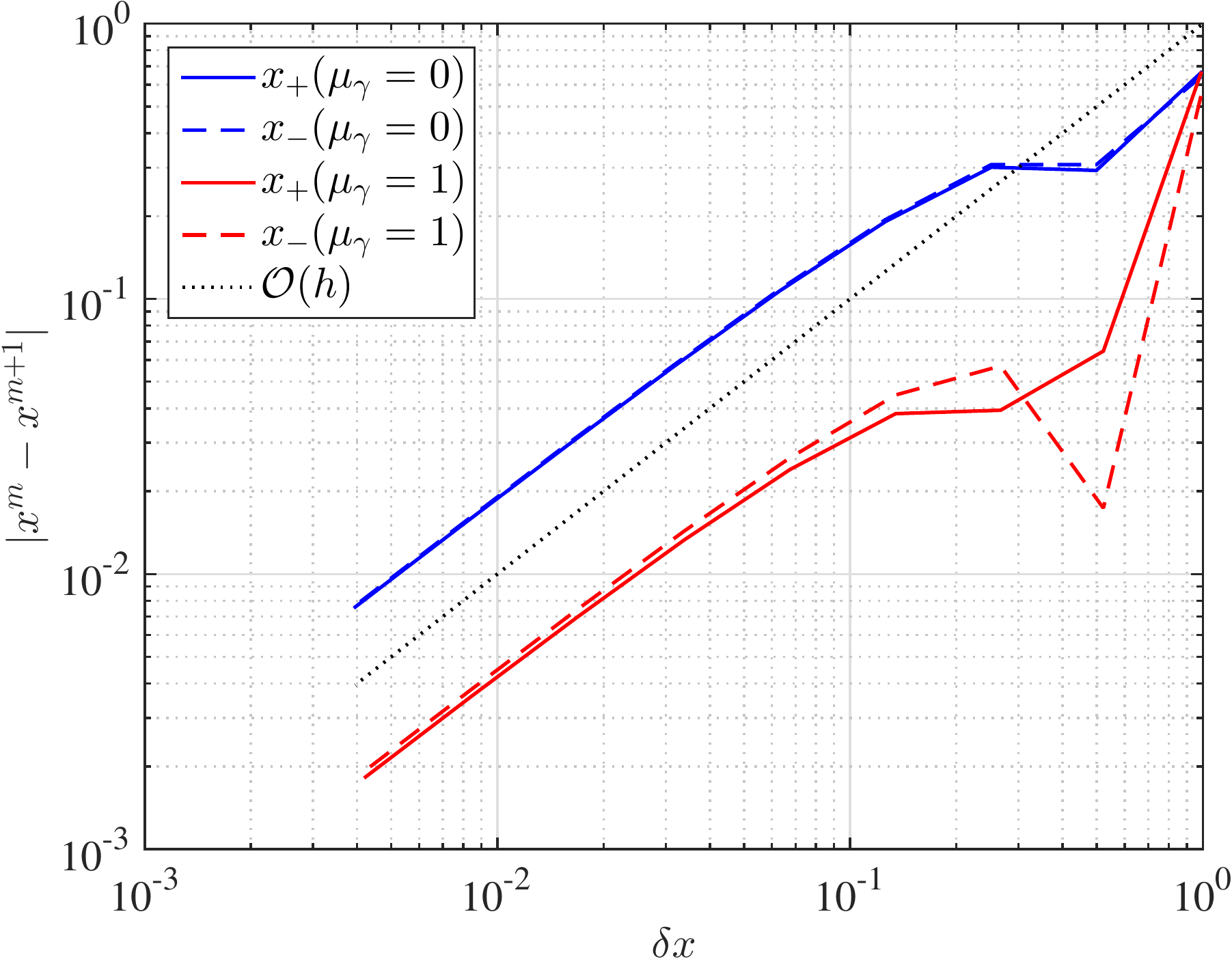}\,%
\includegraphics[height=0.26\textwidth]{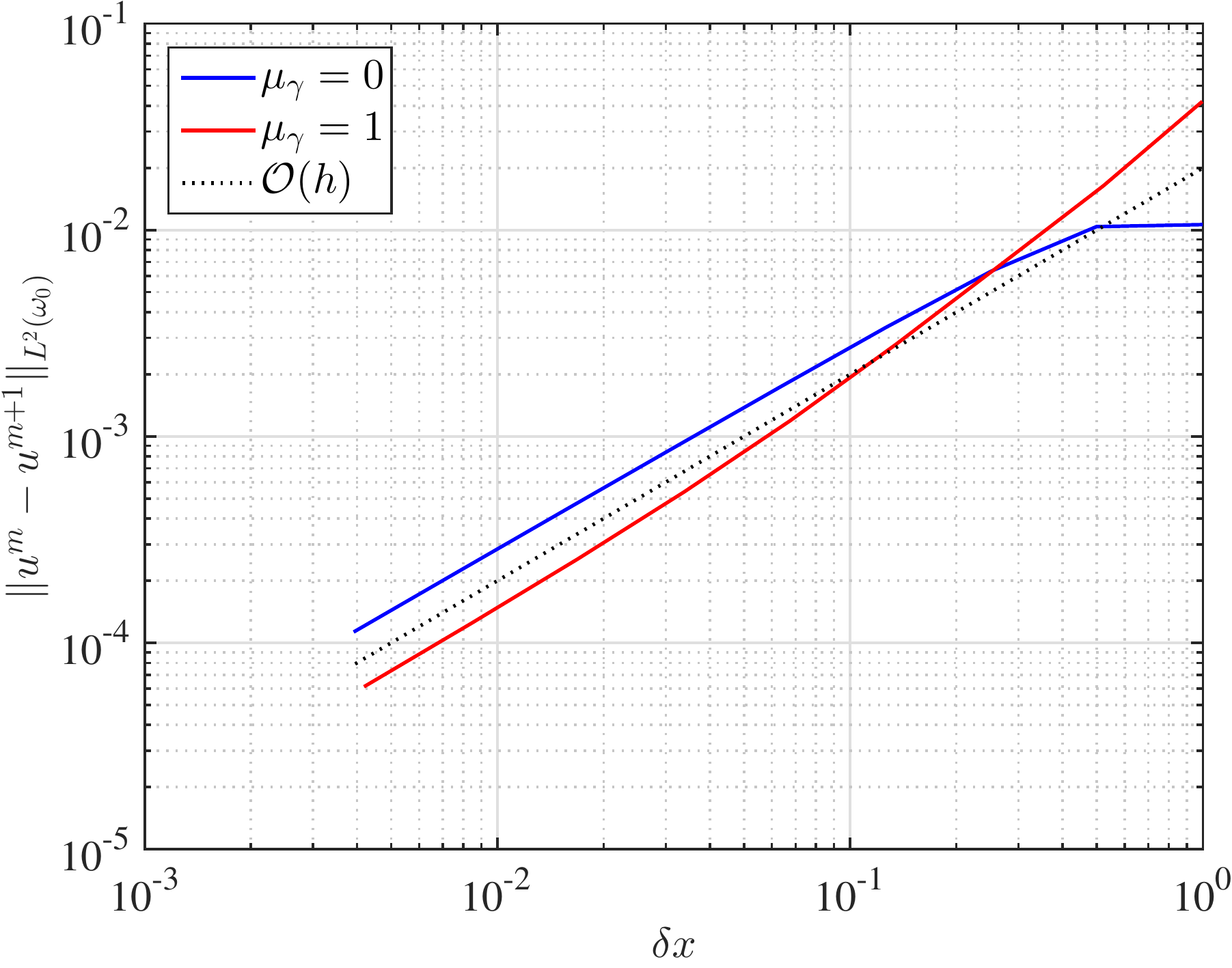}\,%
\includegraphics[height=0.26\textwidth]{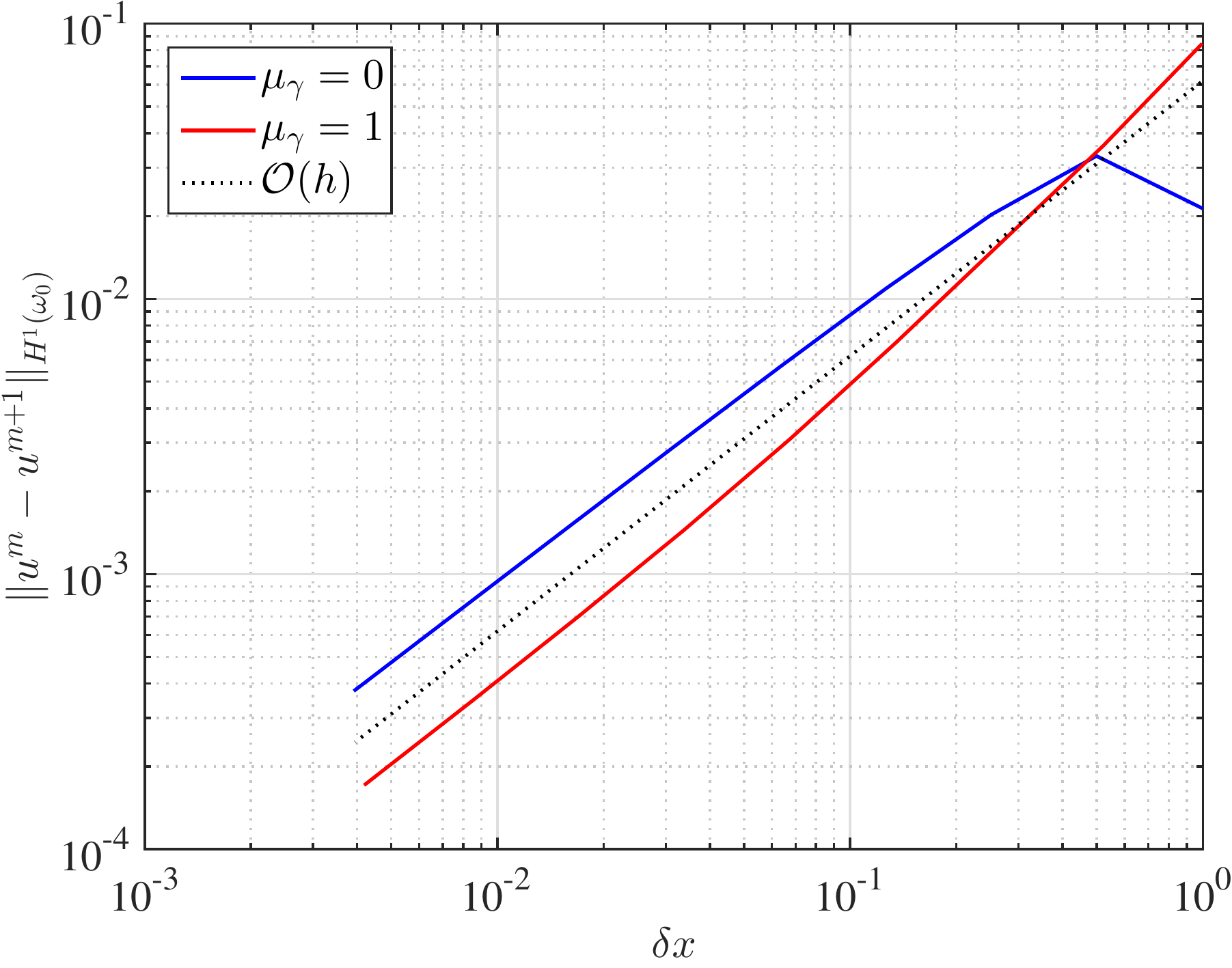}%
\caption{Convergence of solutions for $\mu_\gamma=0$ (blue) and $\mu_\gamma=1$ (red) for \textbf{(left)} the positions of the contact points at $t=8$ and \textbf{(middle)} for the $L^2$ norm of the solution and \textbf{(right)} for the $H^1$ norm of the solution}
\label{fig:convergence}
\end{figure*}

Finally, the corresponding contact line velocities $\dot{x}_\pm$ for $\mu_\gamma=0,1$ are shown in the right panel of Fig.~\ref{fig:solution}. As in the middle panel, the equality of velocities $\dot{x}_-\approx \dot{x}_+$ for $t>3$ suggests that the solution is close to a traveling wave. As expected, with added contact line dissipation the traveling wave speed for $\mu_\gamma=1$ is slower than the speed for $\mu_\gamma=0$. Additionally, for 
both $\dot{x}_\pm$
we also observe that the evolution obeys the predicted contact line law \eqref{eqn:contactlinelaw}, \emph{i.e.},
\begin{align}
\dot{x}_\pm = \pm\mu_\gamma^{-1}\left(\frac{1}{2}|\partial_x h\bigl(t,x_\pm(t)\bigr)|^2-1\right),
\end{align}
which is visible in the overlap of the dotted curve and the red dashed curve/red full curve for $\mu_\gamma=1$.

\begin{figure}[t]
\centering
\includegraphics[width=0.4\textwidth]{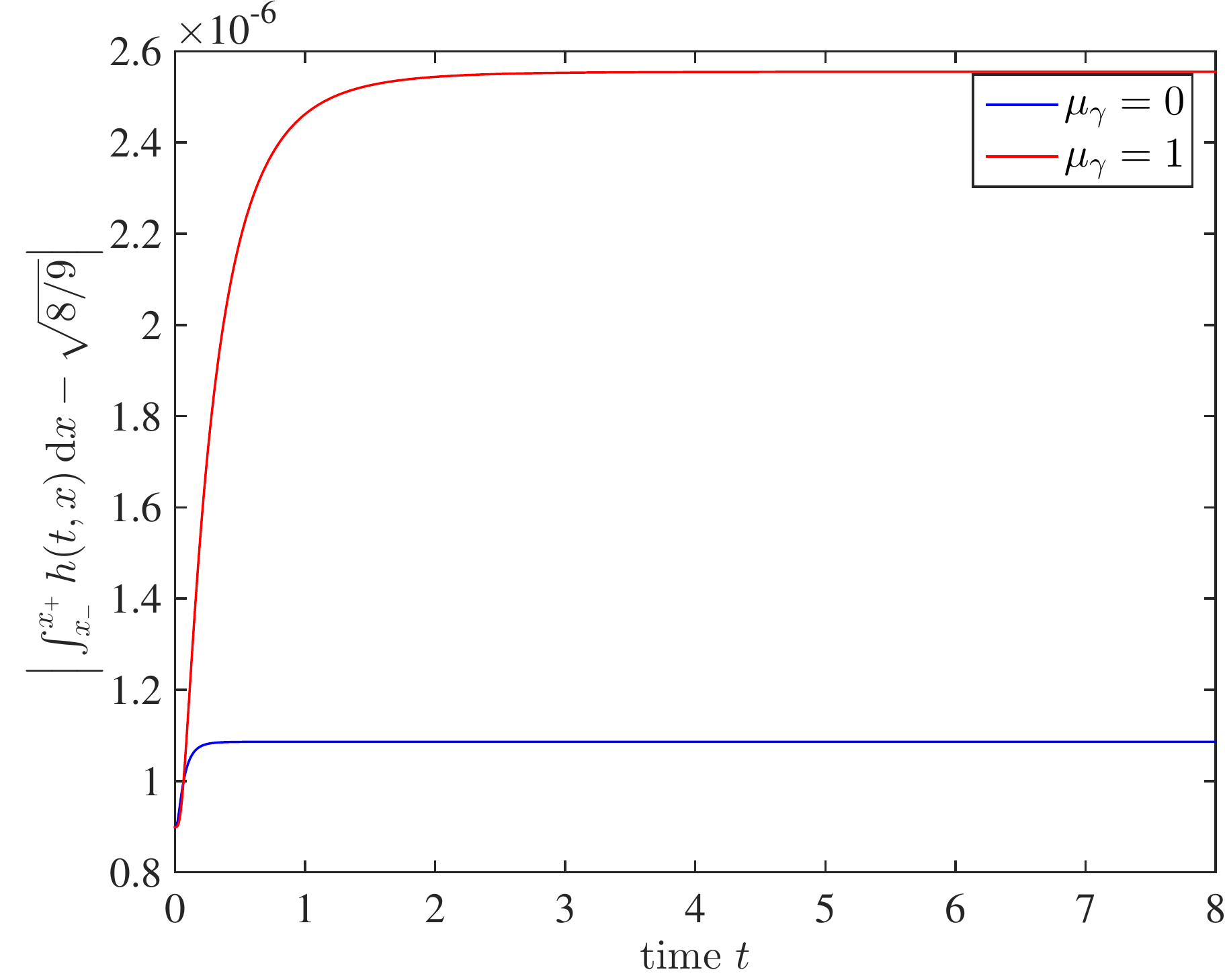}\,%
\caption{Deviation of volume from exact value $\int h{\rm d}x=\sqrt{8/9}$ for solution with $\mu_\gamma=0,1$ for spatial discretization level $m=10$ as a function of time}
\label{fig:volumes}
\end{figure}

In order to study the experimental convergence order of solutions while $\delta x\sim 2^{-m}L\to 0$, we compare solutions at level $m$ with solutions at neighboring levels $m+1$ at fixed time $T=8$. The left panel of Fig.~\ref{fig:convergence} shows the linear convergence of $|x_\pm^m-x_\pm^{m+1}|$ as $\delta x\to 0$ in both cases $\mu_\gamma=0,1$ with similar errors for $x_-$ and $x_+$. However, note that the magnitude of the error appears slightly better for $\mu_\gamma=1$.
The middle and right panel of Fig.~\ref{fig:convergence} show the convergence of the height profile $h(t,\psi^m_0(t,x))$ with $\psi^m_0(t,\cdot):\omega_0\to\omega(t)$ on a fixed domain $\omega_0$, since solutions for different $m$ are generally defined on different domains. Note that we have a   linear rate of convergence in the $L^2(\omega_0)$ and even in the $H^1(\omega_0)$ norm. Additionally, in Fig.~\ref{fig:volumes} we show that the deviation of the total volume during the evolution is of the order $10^{-6}$.

This shows that this type of algorithm is able to reproduce the predicted contact angle dynamics rather accurately and stable. Due to the inherent coupling of space and time in the free boundary problem, it is certainly a challenging but nevertheless interesting question how to construct a  higher order method.

\section*{Conclusion}
This paper presented a gradient flow approach to contact line dynamics that is based on a quadratic dissipation mechanism at the contact line. The original model is constructed for the Stokes flow and then reduced to a thin-film model. For the reduced thin-film model the variational form including constraints is stated explicitly and a time- and space-discretization is proposed. In one spatial dimension a novel numerical scheme is constructed explicitly and used to study the motion of gravity driven droplets towards traveling wave solutions. These solutions confirm the general expectation of advancing and receding contact angles, where the asymmetry of moving droplets can be
strongly affected by the contact line dissipation. It is expected that such a modification of the thin-film dynamics has a strong impact on pattern formation processes observed during dewetting in the physically relevant case for $\Omega(t)\subset\mathbb{R}^3$ corresponding to $d=2$.

\section*{Acknowledgement}
\noindent
I am thankful for discussions with Luca Heltai (SISSA) and Marita Thomas (WIAS). This research is carried out in the framework of \textsc{Matheon} supported by the Einstein Foundation Berlin.

\appendix
\section{Appendix}
\label{sec:appendix}
\subsection{Hints concerning the notation}
\label{appendix:hints}
The integration $\int\ldots{\rm d}\bx\,{\rm d}z$, $\int\ldots{\rm d}\bx$,  $\int\ldots {\rm d}s$, $\int\ldots {\rm d}\gamma$ refers to the $d+1$, $d$,
$d$, and $d-1$ dimensional integration over $\Omega$, $\omega$, $\Gamma$, and $\gamma$, respectively.
%
For $d=1$ the latter is the point evaluation at $(x_\pm,0)\in\mathbb{R}^2$.
%
%
The signed mean curvature is defined
\begin{align*}
\kappa=\frac{1}{d}\sum_{i=1}^{d} \kappa_i,
\end{align*}
as the mean of the principle curvatures $\kappa_i$. The discretization of $\kappa$ uses the identity $\kappa\bnu_\Omega=\underline{\Delta}{\rm id_{\Gamma}}$ for $\Gamma\subset\partial\Omega$, where details concerning the definition of the Laplace-Beltrami operator using the coordinate identity vector field ${\rm id_\Gamma}(\mathbf{x},z)=(\mathbf{x},z)$ can be found in
 \cite{bansch2001finite,dziuk1990algorithm}.
%

\subsection{ALE mapping of $\dot{h}$}
\label{appendix:ALE}
In one dimension $d=1$ consider the reference interval at time $t_0$ given by  $\omega_0=\bigl(x^0_-,x^0_+\bigr)$ with $x^0_\pm = x_\pm(t_0)$. For arbitrary time $t$ we define $\boldsymbol{\xi}(t,\cdot):\omega_0\to\omega(t)$ as
\begin{align*}
\boldsymbol{\xi}(t,y)= \Big(x_+(t)-x_-(t)\Big)\frac{y-x^0_-}{x^0_+-x^0_-}+x_-(t),
\end{align*}
so that $\boldsymbol{\xi}(t_0,\cdot)={\rm id}_{\omega_0}$. Let
$h(t,\cdot):\omega(t)\to\mathbb{R}^+$ non-negative be defined on a time-dependent domain $\omega(t)$ with $h(t,\cdot)\equiv 0$ on the boundary $\partial\omega(t)$. The pull-back $H(t,\cdot):\omega_0\to\mathbb{R}^+$ of $h(t,\cdot)$ to the reference domain $\omega_0$  is defined
\begin{align*}
H(t,y)=h\bigl(t,\boldsymbol{\xi}(t,y)\bigr),
\end{align*}
with the corresponding time-derivative
\begin{align*}
\dot{H}(t,y)=\dot{h}(t,\boldsymbol{\xi}) + \dot{\boldsymbol{\xi}}\cdot\nabla h(t,\boldsymbol{\xi}).
\end{align*}
On $\partial\omega_0$ we have $\dot{H}\equiv 0$ and thereby $\dot{h} + \dot{\boldsymbol{\xi}}\cdot\nabla h=0$
with $\dot{\boldsymbol{\xi}}=\dot{x}_\pm$.
Since $\bnu=-\nabla h/|\nabla h|$ defines the outer normal, we can
reconstruct the normal component of $\dot{\boldsymbol{\xi}}$ as
\begin{align*}
\dot{\boldsymbol{\xi}}\cdot\bnu=\frac{\dot{h}}{|\nabla h|}.
\end{align*}
%
Except for the explicit form of $\boldsymbol{\xi}$ using $x_\pm$, all steps can be generalized to higher dimensions by solving an additional interpolation problem for $\boldsymbol{\xi}$.

\subsection{Uncompensated Young force}
\label{appendix:young}
The uncompensated Young force $\mathbf{f}_\gamma=(\theta_\ell\bnu_{\Gamma_\ell} +
 \theta_\text{s$\ell$}\bnu_{\Gamma_\text{s$\ell$}})$ in the contact line model \eqref{eqn:clm} is multiplied with vectors $\mathbf{t}$ parallel to the $\bx$-plane. When using $\mathbf{t}=\bnu_{\Gamma_\ell}=(\bnu,0)$, with $\bnu$
 the outer normal on $\partial\omega$, this produces
 \begin{align}
\label{eqn:clm1}
 \mu_\gamma \bnu\cdot\bu_\bx = \theta_\text{s$\ell$}-\theta_\ell\cos\vartheta,
 \end{align}
 with surface tensions $\theta_\text{s$\ell$}=\theta_\text{solid,liquid}-\theta_\text{solid,air}$ and $\theta_\ell=\theta_\text{liquid,air}$. The equilibrium contact angle, when it can be defined, is $\theta_\text{s$\ell$}=\theta_\ell\cos\vartheta_\text{e}$ so that \eqref{eqn:clm1} can be written
 \begin{align}
\label{eqn:clm1a}
 \bnu\cdot\bu_\bx = \frac{\theta_\ell}{ \mu_\gamma}\bigl(\cos\vartheta_\text{e}-\cos\vartheta\bigr).
 \end{align}


\bibliographystyle{unsrt}
\bibliography{si}

\end{document}